\documentclass[10pt,twocolumn,letterpaper]{article}

\usepackage[pagenumbers]{cvpr} % To force page numbers, e.g. for an arXiv version

\usepackage{graphicx}
\usepackage{amsmath}
\usepackage{amssymb}
\usepackage{booktabs}
\usepackage{listings}
\usepackage{todonotes}
\usepackage[pagebackref,breaklinks,colorlinks]{hyperref}

\usepackage[capitalize]{cleveref}
\crefname{section}{Sec.}{Secs.}
\Crefname{section}{Section}{Sections}
\Crefname{table}{Table}{Tables}
\crefname{table}{Tab.}{Tabs.}

\def\confName{CS 7643, Georgia Tech}
\def\confYear{Fall 2023}

\begin{document}

% \listoftodos

\title{Physics Informed Neural Network for Option Pricing}

\author{Ashish Dhiman\\
{\tt\small ashish1610dhiman@gmail.com}
\and
Yibei Hu\\
{\tt\small yibei.careers@gmail.com}
}
\maketitle

%%%%%%%%% ABSTRACT
\begin{abstract}
   We apply a physics-informed deep-learning approach \footnote{Code available at: \url{https://github.com/ashish1610dhiman/pinn_option_pricing}} to the Black-Scholes equation for pricing American and European options. We test our approach on both simulated as well as real market data, compare it to analytical/numerical benchmarks. Our model is able to accurately capture the price behavior on simulation data, while also exhibiting reasonable performance for market data (with an improvement of \~30\% over benchmark). We also experiment with the architecture and learning process of our PINN model to provide more understanding of convergence and stability issues that impact performance. 
\end{abstract}

% \todo[inline]{Correct Author and stuff: anyone}
% \todo[inline]{Explain Abstract: Anshit}

%%%%%%%%% BODY TEXT
\section{Introduction}
Today, finance dramatically benefits from the introduction of Artificial Intelligence and Deep Learning methodologies. Among the several topics discussed in the last years, we deal with numerical methods for option pricing. Options are financial derivative contracts that give the buyer the right to buy or sell an underlying asset at a fixed price within a specified period or on a certain expiration date. The most important financial option styles are European and American. Call option grants the right to buy, and the Put option grants the right to sell \cite{option_pricing_models}. European guarantees the buyer to purchase/sell the asset at the exercise price only at the maturity date. However, the American options allow the buyer to purchase/sell the asset prior to and including the maturity date \cite{hull_book}. European options look only to maturity price, while American ones also consider price dynamics from the contract's start date to maturity.

\subsection{Option pricing}
An option price consists of intrinsic value (a measure of the profitability of an option) and time value (which is based on the expected volatility of the underlying asset and the time left until the option expiration date). Determining the fair value of option prices depends on several factors, including asset price, strike price, time until expiration, volatility, and the risk-free interest rate. %The more time remains until the option expires, the greater the time value of the option is.

The primary goal of option pricing is to work out the probability of whether the option is “in-the-money” or “out-of-money” when it is exercised. Option pricing is crucial for traders, investors, and financial institutions in making informed decisions about buying, selling, or hedging risks against certain underlying assets. Precise estimation of the option price helps stabilize the financial market, as financial portfolios and strategies are adjusted according to the changes in the option price \cite{pricingModelPerformance}. The problem of robust option pricing becomes even more pressing in the current times, with the volatile market conditions and unforeseen changes in Treasury Yield curves.

Partial Differential Equations (PDE) method can be applied to option pricing problems. That is, to compute the price function as the solution of a PDE. One such method is the framework of the Black–Scholes model \cite{bs_model}, where a parabolic nonlinear PDE is used to describe option price dynamics. Furthermore, specific initial and boundary conditions are exploited to represent some contract features mathematically, e.g. call or put, European or American. The BS model \ref{eq:bs_pde} has been the industry bedrock in estimating the fair value of options, with numerous modifications proposed \cite{numericEquation}.

\begin{equation}
f:= \frac{\partial V}{\partial t} + \frac{1}{2} \sigma^2 S^2 \frac{\partial^2 V}{\partial S^2} + rS \frac{\partial V}{\partial S} - rV = 0
\label{eq:bs_pde}
\end{equation}

where $V$ is the option price, $t$ is time, $\sigma$ is the volatility of the underlying asset, $S$ is the spot price of the underlying asset, and $r$ is the risk-free interest rate.

While the Black-Scholes equation (\ref{eq:bs_pde}) can be analytically solved for European options, the introduction of American options, which allows for the exercise of the contract at any point within the specified time period, makes the analytical solution infeasible. Consequently, numerical techniques such as Finite Difference and Finite Element are used to approximate the option price for American options.

Our physics-informed solution is driven by data, and the above PDE (\ref{eq:bs_pde}) will compute the hidden state (or Price) $V(t,S)$ of the system (or option), given boundary data, measurements and fixed model parameters (risk free rate, $r$) and volatility ($\sigma$):
\begin{equation}
V(s,t) + N[V(s,t)] = 0, \quad  \forall (S,t) \in \Omega\\
\end{equation}

where $N$ is the BS nonlinear differential operator corresponding to the BS eqn on our \textbf{input domain ($\Omega$) of spot price and time}, with $V(s,t)$ being approximated by a NN.

\subsection{Deep Neural Networks}
The increase use of neural networks as function approximators \cite{HORNIK1989359} and numerical solvers provides a new frontier for machine learning and numerical analysis.

In Deep Neural Networks (DNNs), automatic differentiation plays a pivotal role. Training a neural network involves providing the loss function. To comprehend the inner workings of a DNN, understanding a single perceptron's functionality is key. This fundamental unit, constituting a single neuron and activation function, forms the crux of a DNN. While only one layer can be used to make predictions, researchers often opt to make their neural networks “deep” to improve the flexibility and modeling capabilities of the network. Neural networks are composed of multiple layers, and higher layers identify complex patterns crucial for accurate predictions.

\subsection{Our Contribution}
The recent Deep Learning strategy, namely the Physics Informed Neural Networks (PINNs) \cite{raissi2019physics}, provides a way to complete the automatic differentiation step, and provide the numerical resolution of PDE. In this paper, we applied this novel Deep Learning strategy: PINN to solve option pricing problem. The approach takes the same inputs as required for the BS model i.e Spot price and time. We implemented the following throughout our project:
\begin{itemize}
    \item We applied a cutting-edge framework, namely the PINNs, to solve the American and European put option pricing problem from a numerical perspective. Earlier studies that have been conducted are applied exclusively to either one of the products, while our exercise covers both.
    \item Unlike \cite{pinn_option_pricing}, which compared the American option solutions to the numerical solutions, we compared them to the market prices which will give us a better benchmark for approximation.
    \item We also conducted experiments with varying architectures such as RNN-based architecture of PINNs, applying in-depth knowledge in the computational finance domain to alter earlier architecture.
\end{itemize}

We believe we would be able to significantly contribute to the option pricing domain through our project. The applications of PINNs for Finance are, to the best of our knowledge, still little studied and only restricted to standard European option pricing. There are few works which address this problem. Our project can help researchers to explore the usage of neural networks in traditional pricing domain. 

% \todo[inline]{Format and explain above BS eqn: Pratyusha}

\section{Related Works}
% \todo[inline]{Anshit}
%% Copy references from -> https://math.dartmouth.edu/theses/undergrad/2021/Louskos-thesis.pdf

Traditionally methods like finite difference and finite element methods \cite{wrds, blackscholes_equation, numericEquation, lamberton2018binomial}, have tackled PDEs and options pricing, each with its limitations. Recent years have seen a surge in alternative techniques for solving differential equations, such as Lie symmetry\cite{wrds}, B\"{a}cklund transformation \cite{bs_model}, and neural networks \cite{moseley2020solving, RAISSI2019686, Sirignano_2018}. Neural networks, especially Physics-Informed Neural Networks (PINNs) \cite{moseley2020solving, RAISSI2019686, Sirignano_2018}, have shown promise in solving complex PDE problems with success in various physics applications.

In finance, PINNs have shown potential primarily in European option pricing \cite{pinn_option_pricing,wang2023experts}. Some studies have applied PINNs to solve European option pricing problems under the Black-Scholes and Heston models \cite{pricingModelPerformance, HORNIK1989359, pinn_option_pricing, wang2023experts}. They address the American call option pricing as a European one when dividends are absent.

\section{Methodology}
% \todo[inline]{Peter}
\subsection{Physics Informed Neural Network (PINN)}

A PINN (Physics-Informed Neural Network) is a type of machine learning model that combines deep neural networks with physical principles to solve partial differential equations (PDEs) and other physics-based problems \cite{RAISSI2019686}. By formulating PDEs, boundary conditions, or other physics-based constraints as components of the neural network's objective function, PINNs can simultaneously learn from data and respect the governing equations. This distinctive combination of data-driven learning and physics-based modeling makes PINNs a powerful tool for solving a wide range of problems, even in scenarios where traditional numerical methods may struggle or when data is scarce or noisy \cite{Sirignano_2018}. PINNs have been successfully applied to various use cases such as wave equation \cite{moseley2020solving}, etc. In this approach, the neural network is trained to minimize a combined loss function, which includes both data-driven terms to fit observed information and physics-driven terms to enforce the underlying physical laws. PINNs are proven to be more reliable than alternative solutions by leveraging the accuracy, scalability and computational efficiency of data-driven supervised neural networks while using physics equations given to the model to encourage consistency with the known physics of the system and thereby reduce the convergence issues and extrapolate accuracy beyond the available data.

\subsection{PINN for Option Pricing}
\label{sec:pinn_option_pricing}
% \todo[inline]{Add diff b/w european and american: Peter}
To apply the PINN approach to Option pricing, we need to define the initial and boundary values for the Black Scholes equation. The PDE for American options, which allow for early exercise at any time within the specified period, is notably more intricate than the PDE for European options. Specifically, American options involve a free-boundary condition, where the boundary condition varies with time. This is detailed more in next section.

After defining the boundary/initial values the next step is then to train our model to minimize the following loss function:
\begin{equation}
    Loss(\beta) =  MSE_{ivp} + MSE_{bvp} + \beta * MSE_{pde}
    \label{eq:pinn_option_loss}
\end{equation}

\begin{equation}
\begin{aligned}
    MSE_{bvp} = \frac{1}{|\Gamma|} \sum_{x_i \in \Gamma} (V(x_i)-\hat{V}(x_i))^2\\
    MSE_{ivp} = \frac{1}{|\Phi|} \sum_{x_i \in \Phi} (V(x_i)-\hat{V}(x_i)^2\\
    MSE_{pde} = \frac{1}{|\Omega'|} \sum_{x_i \in \Omega'} (f(x_i)-0)^2\\
\end{aligned}    
\end{equation}

where $\beta$ is strength of PDE penalty, $\Gamma$ is set of boundary value points, $\Phi$ is set of Initial value points, $\Omega'$ is set of sampled points from complete sample space, $f$ is Black Scholes PDE \ref{eq:bs_pde}. 

The losses $MSE_{bvp/ivp}$ are calculated for boundary value and initial value conditions for the PDE. While $MSE_{pde}$ is the differential equation loss. Here we are trying to minimize the PDE by calculating gradients and forming the PDE itself. We expect that by minimizing loss \ref{eq:pinn_option_loss}, our neural network will be able to accurately predict the price of European and American options, as well as the corresponding Greeks.

% \todo[inline]{Add BVP/IVP for european and american option: Ashish}
\subsubsection{European Call Option}
As for an European call option, we use Black-Scholes formula that involves only the first derivative w.r.t ( with respect to ) time, but the first and second derivatives w.r.t the asset value S. We will need one boundary condition w.r.t time (the value at the expiry time t=T) and two boundary conditions with respect to S (based on the behavior of the option at S=0 and as S $\rightarrow \infty$).

\begin{itemize}
  \item BVP1 for $S = S_{\min}$: $V = 0$
  \item BVP2 for $(S = S_{\max})$: $ V = S_{\max} - K \cdot e^{-rt}$
  \footnote{K: strike price}
  \item IVP for $t=T$: $V = \max(S-K,0)$
\end{itemize}

The expiry time condition for the call option is C(T,S)= max(S-K,0) as the initial value condition at its expiration time T for all S. If the underlying asset becomes worthless, then it will remain worthless, so the option will also be worthless.\cite{blackscholes_equation}. On the other hand, if S becomes very large then the option will almost certainly be exercised, and the exercise price is negligible compared to S. Thus the option will have essentially the same value as the underlying asset itself. The benchmark used for european call is:

The true solution for the European call option, as given by the Black-Scholes formula, is expressed by the equation:
\begin{equation}
C(S, t) = S \cdot N(d_1) - X \cdot e^{-r(T-t)} \cdot N(d_2)
\end{equation}
where:
\begin{align*}
&d_1 \text{ and } d_2 \text{ are calculated as follows:} \\
&\quad d_1 = \frac{\ln(S/X) + (r + \sigma^2/2)(T-t)}{\sigma \sqrt{T-t}} \\
&\quad d_2 = d_1 - \sigma \sqrt{T-t} \\
\end{align*}

\subsubsection{American Put Option}
To price an American put option, one needs to consider the optimal strategy of exercising the option before expiry, which depends on the underlying asset's price, time, and other factors like volatility and interest rates. We look for a function $V = V_{am}(T, s)$ and a boundary $S_{f}(t)$ with $0<= S_{f}(t)<=K=S_{f}(T)$ such that, for all S, t, we have

\begin{equation}
\frac{\partial V}{\partial t} + \frac{1}{2} \sigma^2 S^2 \frac{\partial^2 V}{\partial S^2} + rS \frac{\partial V}{\partial S} - rV <= 0,
\label{eq:am_fe1}
\end{equation}
and
\begin{equation}
V_{Am} (t, S) >= max(K-S,0).
\label{eq:am_fe2}
\end{equation}

Moreover, at least one of these two inequalities (6),(7) is always an equality. More precisely considering with the boundary conditions based on the asset price:

(a) if $S <= S_{f}(t)$ then the option should be exercised immediately and $V_{Am}(t, S)=max(K-S, 0) = K-S$;

(b) if $S > S_{f}(t)$ then the option should be held for the time being, and $V_{Am}(t,S)>max(K-S, 0)$. Hence we have the following boundary conditions:

% The payoff from immediate exercise will be small, but it is possible that the asset price will fall, giving a much larger payoff from exercising the option at some future date (even allowing for discounting). In this situation, the option value if we continue to hold it is more than will be obtained from immediate exercise.

\begin{equation}
(\frac{\partial V}{\partial t} + \frac{1}{2} \sigma^2 S^2 \frac{\partial^2 V}{\partial S^2} + rS \frac{\partial V}{\partial S} - rV)*(\max(K-S,0) - V) = 0
\label{eq:am_fe_0}
\end{equation}

\begin{itemize}
  \item BVP1 for $S = S_{\min}$: $V = K$
  \item BVP2 for $(S = S_{\max})$: $ V = 0$
  \item IVP for $t=T$: $V = \max(K-S,0)$
\end{itemize}

where K is the strike price, $V_{Am}$ is the value of the American put option value as a function of time t and the underlying asset price S. $S_{f}(t)$ is the early exercise boundary, which is a function of time that satisfies $0<=S_{f}(t)<=K$. The exercise boundary is the threshold at which it becomes optimal for the holder to exercise the American options.

% \begin{center}
% \begin{tabular}{ |c|c|} 
% \hline
% Option Types & Optimal Conditions to exercise\\
% \hline
% Put Option & Asset's price exceeds a certain level\\
% \hline
% Call Option & Asset's price falls below a certain level\\
% \hline
% \end{tabular}
% \end{center}
% We have the boundary condition $V_{Am}(t,S)$ $\rightarrow 0$ as S $\rightarrow \infty$, and from (a) we have $V_{Am}(t,0)$=K and $V_{Am}(T,S)$=Max(K-S,0).

We benchmark our PINN solution for American Put by applying finite difference and Binomial model on the American option problems. The FDM divides the domain into uniform - possible to be non-uniform - intervals of spaces to make a mesh. 
The binomial model for American put options operates by iteratively evaluating the option's value at each discrete time step, considering the possibility of early exercise at each point. It incorporates a dynamic pricing approach based on the potential future stock price movements, allowing for optimal decision-making regarding the exercise of the put option throughout its life.

% Let h be the mesh width and approximate the derivatives of f by working backward from expiry to the present with terminal and boundary conditions and early exercise at each step as follows:
% \begin{equation}
%     \frac{\partial f}{\partial x_{-}} = \frac{f(x) - f(x - h)}{h}
%     \label{eq:backward method}
% \end{equation}
\section{Dataset}
% \todo[inline]{Prathyusha}
\label{sec:data}

To comprehensively showcase the robustness of our approach, we employ two distinct types of data, each serving specific purposes in the validation and application of our model.

Firstly, simulated data plays a pivotal role in validating the PINN methodology. This dataset is carefully crafted using a uniform grid or mesh spanning both the price and time dimensions. By generating synthetic data, we create a controlled environment that allows us to systematically evaluate the performance of our model, as well as the effect of different training schemes. This synthetic dataset becomes invaluable in assessing the model's accuracy, reliability, and generalization capabilities.

\begin{figure}[ht]
    \centering
    \includegraphics[width=0.45\linewidth]{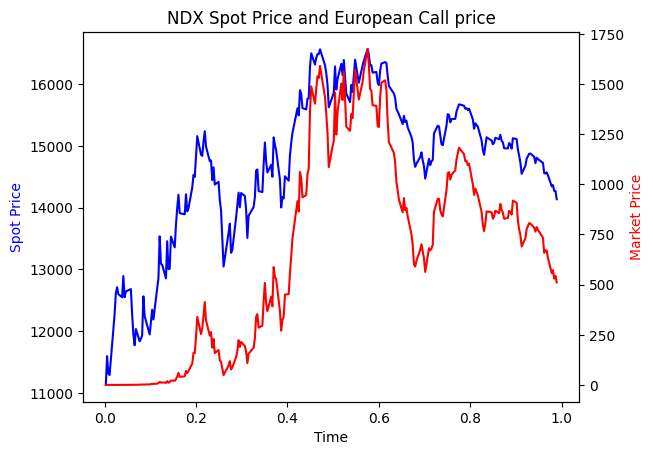}
    \includegraphics[width=0.45\linewidth]{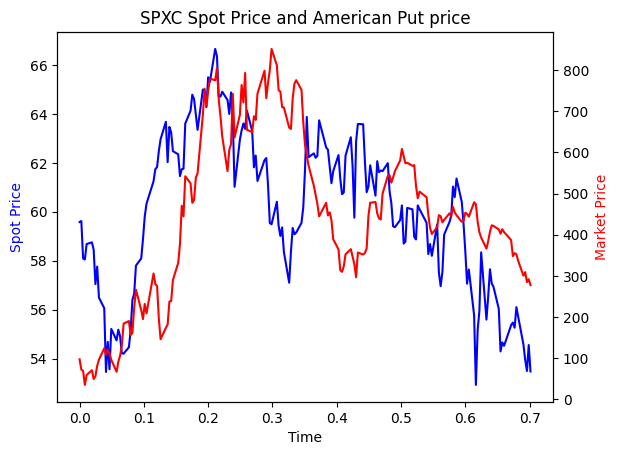}
    \caption{Merged spot price (from Optionmetrics) and market price data (from YF)  for NDX Call (left) and SPCX put (right)}
    \label{fig:merged_data}
\end{figure}

Conversely we enrich our analysis with real-world market data obtained from Optionmetrics through the Wharton Research Data Service \cite{wrds}. This incorporation of authentic market data enhances the practical relevance of our model, showcasing its adaptability and performance in a real-world financial context. This dataset provides us with the trading market price of option instruments. Specifically, we focus on European call options associated with the NASDAQ 100 and American put options linked to SPCX stock. From the said data we specifically get the columns as listed in \ref{tab:key_sources}.

\begin{table}[ht]
  \centering
  \begin{tabular}{lr}
    \toprule
    \textbf{Key} & \textbf{Source} \\
    \midrule
    Bid price & Optionmetrics \\
    Offer price & Optionmetrics \\
    Trading date & Optionmetrics \\
    Strike price (\(K\)) & Optionmetrics \\
    Volatility (\(\sigma\)) & Optionmetrics \\
    Spot price (\(S\)) & Yahoo Finance \\
    \bottomrule
  \end{tabular}
  \caption{Data Key and Sources}
  \label{table:key_sources}
\end{table}

\begin{figure}
    \centering
    \includegraphics[width=0.75\linewidth]{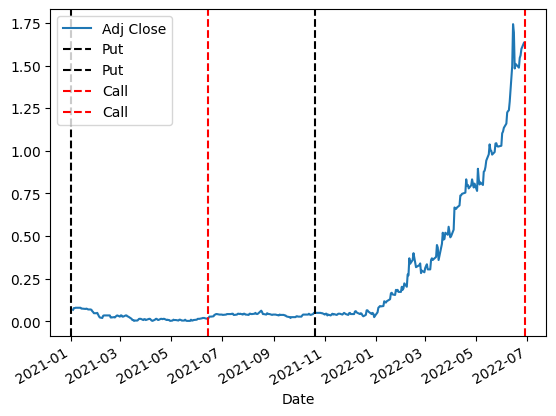}
    \caption{1 year Treasury Yield from Jan 2021 to Jul 2022. Black lines mark the duration of Put option on SPXC, while red line marks the duration of Call option on NDX}
    \label{fig:treasury_yields}
\end{figure}

The input to our PINN model is spot price and current time = (expiry date- trading date). The underlying asset spot prices are extracted from Yahoo Finance and merged with the data extracted from Optionmetrics \ref{fig:merged_data}. We also need the strike price, volatility as the fixed parameters in the \ref{eq:bs_pde} model. While the strike price and volatility are also available in the optionmetrics data. The risk-free rate, a crucial parameter in option pricing models, is derived from the 1-year Treasury yield within the same time frame, which is also taken from Yahoo Finance \ref{fig:treasury_yields}. We compare our PINN model predictions with the  market price. However the OptionMetrics dataset does not directly provide a market price but rather the best bid and ask price, and we calculate the market price with the midpoint of the two.

\begin{equation}
P_{market}  = \frac{P_{bid} + P_{ask}}{2}
\end{equation}

\section{Experiments and Results}

We have conducted experiments with our PINN approach, evaluating its performance on both simulated data and real market data. Our focus has been on applying this technique to analyze two specific financial products: European Call and American Put options. Notably, we have excluded the American Call option from our analysis, as exercising it early is never considered reasonable. As a result the American Call option's price aligns with that of the European Call option due to the impracticality of early exercise. 

In all experiments, our goal is to replicate the Physics-Informed Neural Network (PINN) methodology outlined in \cite{pinn_option_pricing}, although with minor modifications to the neural network (NN) architecture, as relevant to each case. Within each iteration of the training loop, we systematically sample data for three distinct physical conditions associated with the Partial Differential Equation (PDE). Following data sampling, we calculate the loss according to the formulation detailed in \ref{eq:pinn_option_loss}. These individual losses are then aggregated into an unified objective function, which the Neural Network strives to minimize.The sections below detail our results on different problem setups.

\begin{figure}[h]
    \centering
    \includegraphics[width=0.88\linewidth]{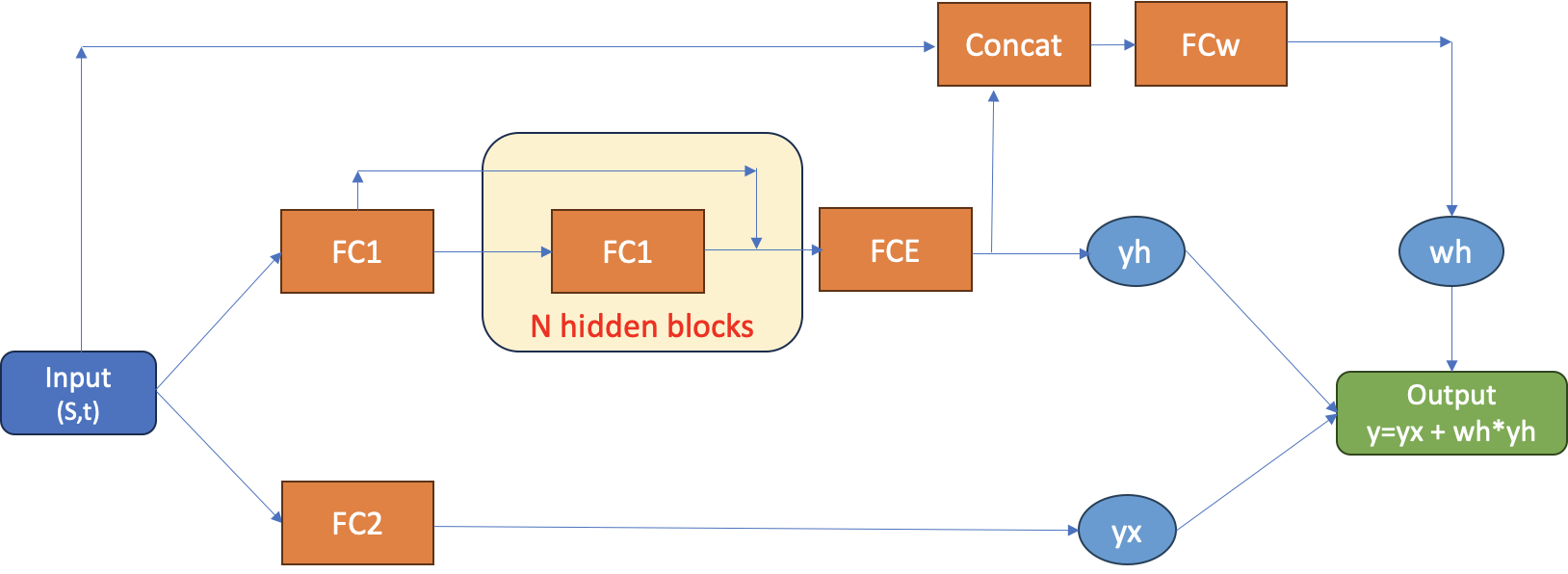}
    \caption{Gated network to capture ramp structure}
    \label{fig:gated_network}
\end{figure}

Taking cues from the intrinsic structure resembling a ramp in the true solutions for both European call and American put, we have devised a network architecture inspired by this pattern \ref{fig:gated_network}. In this NN, the output is a linear combination of two terms. The first term, characterized by minimal hidden layers, captures the simpler aspects of the function. Concurrently, we introduce a second term, equipped with additional hidden layers, aimed at encapsulating the more intricate components of the function.

Our model implementation is informed by best practices gleaned from the existing literature, particularly from Wang et al. \cite{wang2023experts}. We adhere to recommendations such as favoring wider networks over deeper ones, employing a gated structure, integrating residual connections to facilitate gradient flow, and adopting the $\tanh(.)$ activation function, known for its multiple defined gradients. These refinements collectively enhance the network's capacity to effectively capture the complexities inherent in the solution space.

\subsection{Simulated Data}
We generate data for both call and put options using the parameters given in \ref{tab:simulated_parameters}. To validate the accuracy of our European call option results, we compare them against the analytical solution provided in \cite{bs_model}. For the American put option, we establish our benchmark using the Binomial model as outlined in \cite{lamberton2018binomial}. This comparative analysis serves to assess the effectiveness of our generated data against established solutions, providing a robust evaluation of our model's performance.

During each iteration within the training loop, we sample data for the three distinct physical conditions of the PDE. Subsequently, we compute the loss as outlined in \ref{eq:pinn_option_loss}, aggregating them into a unified objective function that the Neural Network aims to minimize.

\begin{figure}[!h]
    \centering
    \includegraphics[width=0.8\linewidth]{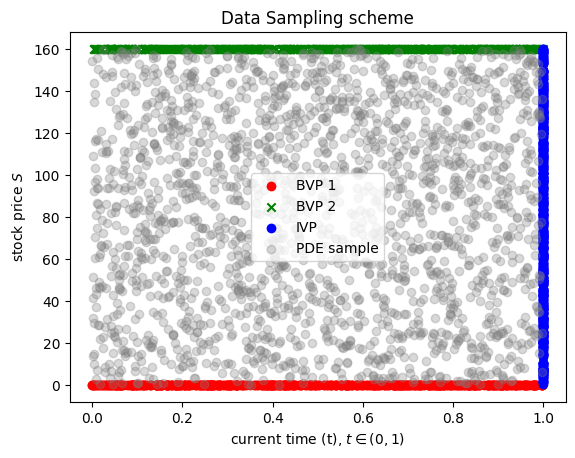}
    \caption{Data Sampling for Simulated data}
    \label{fig:data_sampling}
\end{figure}

\begin{table}[ht]
\centering
\begin{tabular}{|c|c|}
\hline
\textbf{Parameter} & \textbf{Value} \\
\hline
Strike Price, \(K\) & 40 \\
Risk Free rate, \(r\) & 0.05 \\
Volatility, \(\sigma\) & 0.2 \\
\(S_{\text{range}}\) & [0, 160] \\
\(t_{\text{range}}\) & [0, \(1\)] \\
\hline
\end{tabular}
\caption{Simulation Data Parameters}
\label{tab:simulated_parameters}
\end{table}

\subsubsection{Result and analysis on European Call}
Our model undergoes a comprehensive training regimen spanning 25,000 epochs to optimize performance for European call options. The training curve, accessible in \ref{european_call_training_curve}, illustrates a consistent decrease in all losses, including the boundary value losses. Notably, the initial 10,000 epochs exhibit a monotonically decreasing trend, indicating effective convergence. However, in the subsequent epochs, the decrease becomes more oscillatory, hinting at potential convergence challenges. This phenomenon is particularly pronounced for spot prices approaching zero or boundary value 1, where the model tends to oscillate around predicting zero. This observation implies a need for further exploration and refinement to enhance convergence stability in these specific scenarios.

\begin{figure}[!h]
    \centering
    \includegraphics[width=0.88\linewidth]{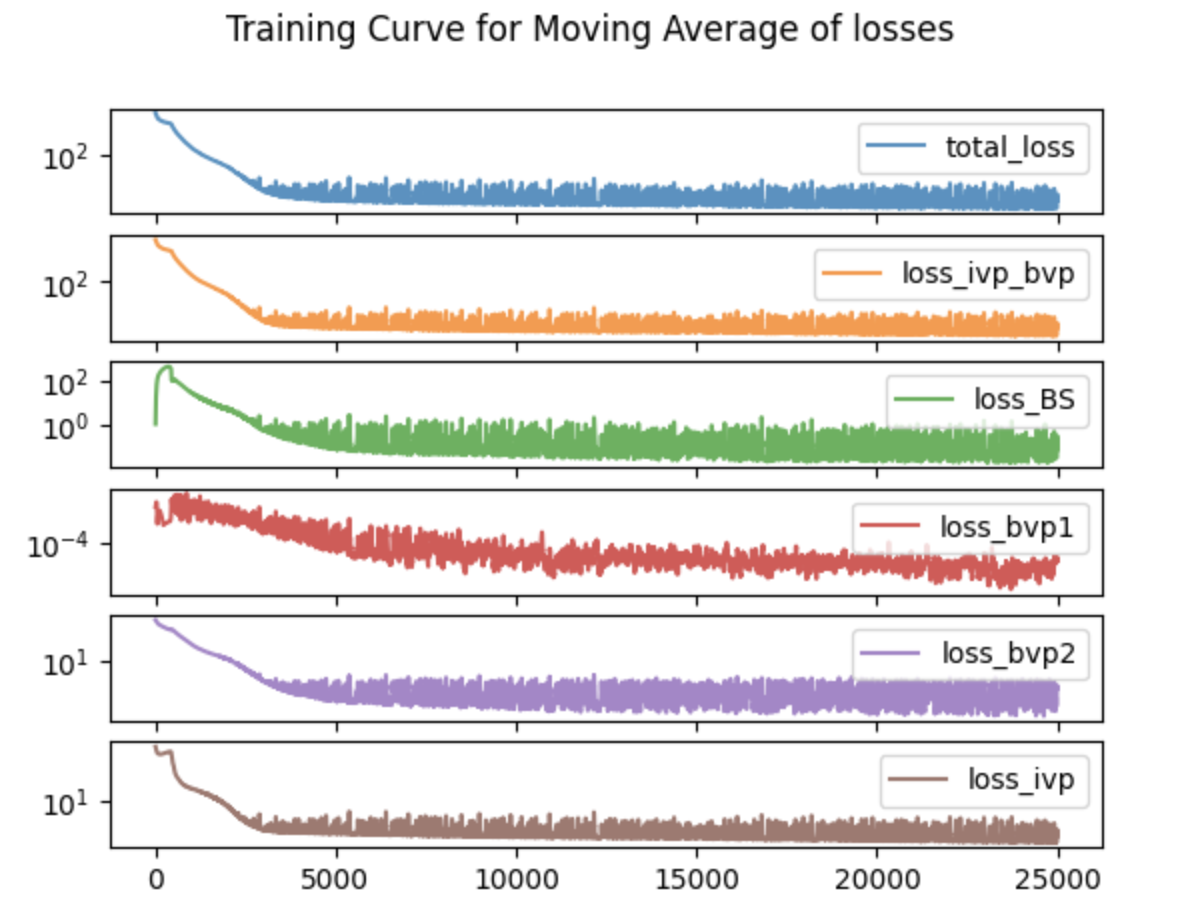}
    \caption{Training Curve for European Call}
    \label{fig:european_call_training_curve}
\end{figure}

\begin{figure}[h]
    \centering
    \includegraphics[width=1\linewidth]{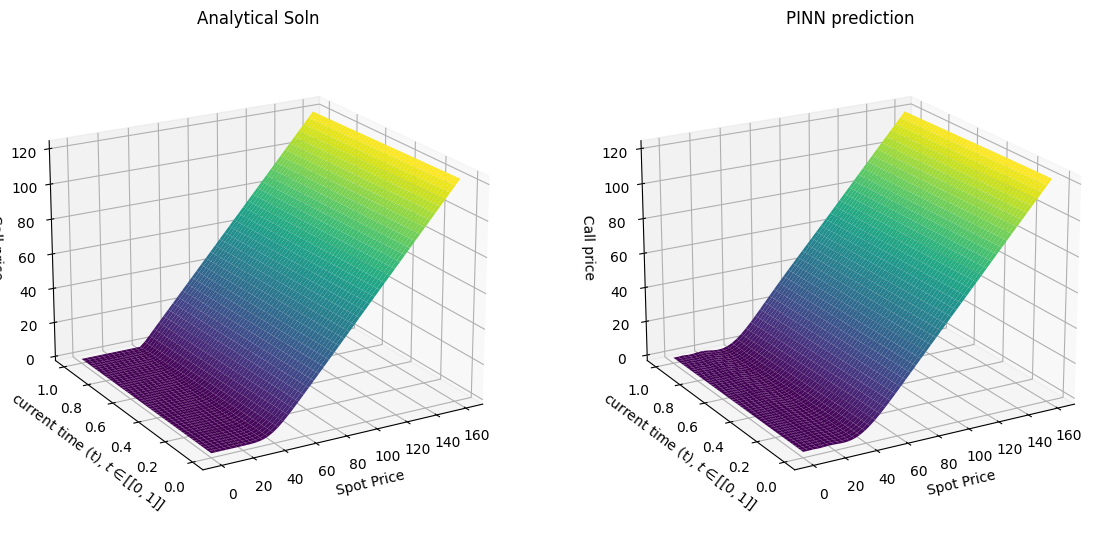}
    \caption{Analytical vs PINN prediction for European Call}
    \label{fig:european_call_results}
\end{figure}

\begin{figure}
    \centering
    \includegraphics[width=0.75\linewidth]{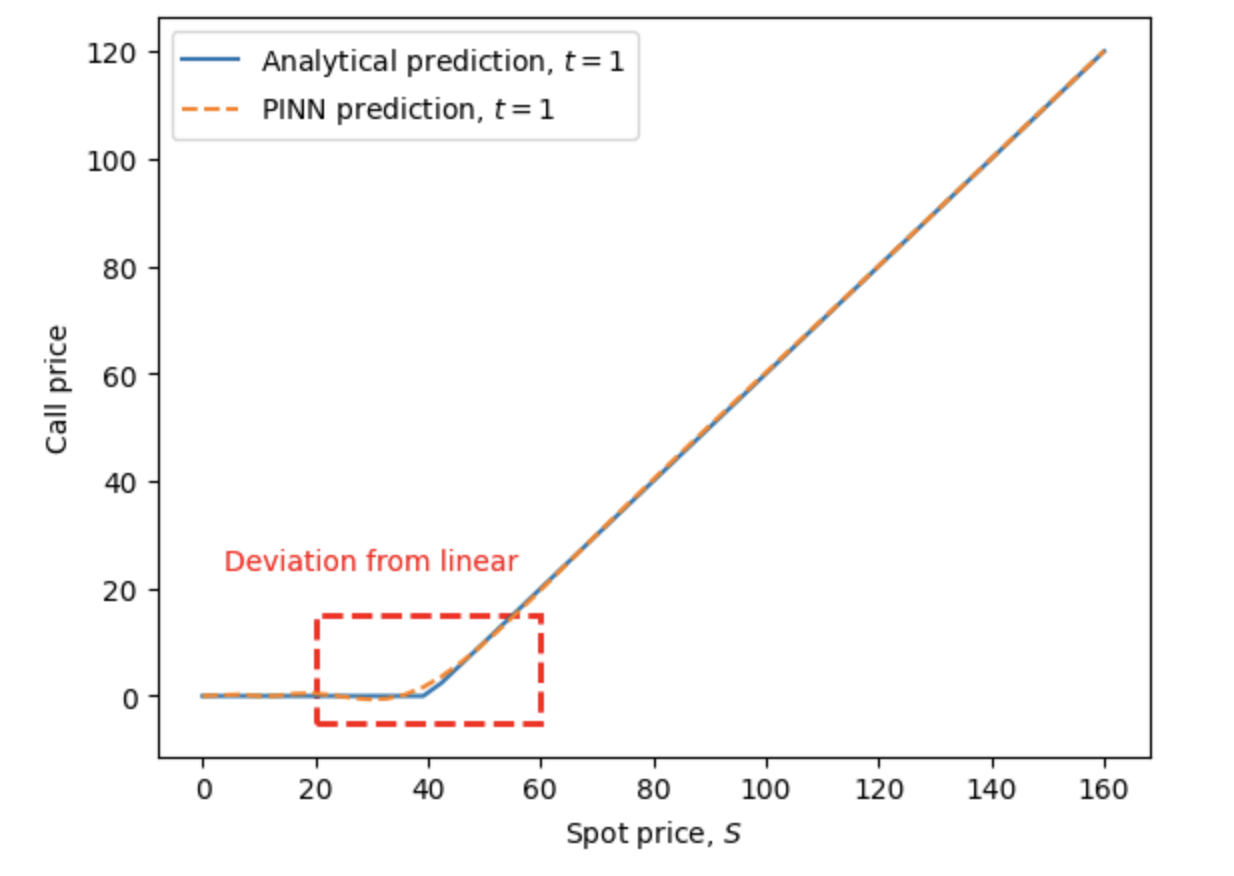}
    \caption{PINN deviates from linear at $t=1$}
    \label{fig:true_vs_pred_non_linear}
\end{figure}

The results from PINN model are also compared to the analytical solution in \ref{fig:european_call_results}, with the performance metrics tabled in \ref{model_performance_metrics}.  Evidently, the PINN model adeptly captures the intricate shape of the European call. However, a nuanced observation arises—specifically, the PINN model tends to exhibit a pronounced nonlinear structure at $t=1$ in lieu of the true linear boundary, as elucidated in \ref{fig:true_vs_pred_non_linear}. Our experimentation delves further into the influence of the parameter $\beta$ Black-Scholes penalty on system dynamics. Notably, augmenting $\beta$ proves conducive to enhancing the neural network's ability to capture non-linearities, albeit occasionally at the expense of generating non-zero predictions for Boundary 1.

\begin{table}[ht]
  \centering
  \begin{tabular}{lcc}
    \toprule
    \textbf{Metric} & \textbf{Eur. Call} & \textbf{Amer. Put} \\
    \midrule
    Total Loss & 0.407 & 0.024 \\
    cor($y_{true}$, $y_{pinn}$) & 1 & 1 \\
    MSE($y_{true}$, $y_{pinn}$) & 0.052 & 0.23 \\
    \bottomrule
  \end{tabular}
  \caption{Model Performance Metrics for European Call and American Put Options}
  \label{model_performance_metrics}
\end{table}

\subsubsection{Result and analysis on American Put}
\begin{figure}[h]
    \centering
    \includegraphics[width=1\linewidth]{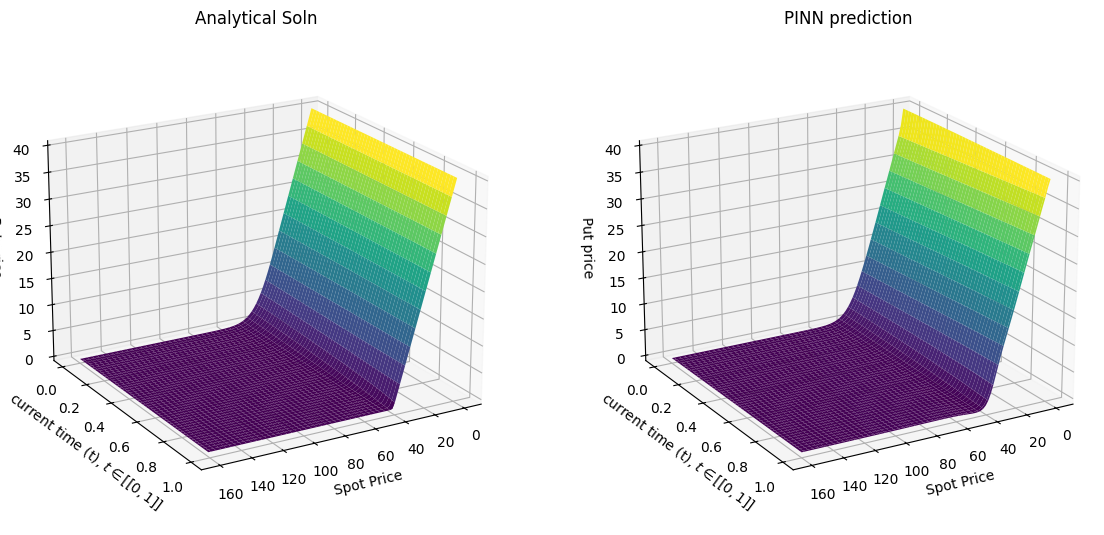}
    \caption{Numerical vs PINN sol'n for American Put}
    \label{fig:american_put_true_pred}
\end{figure}

\begin{figure}
    \centering
    \includegraphics[width=0.45\linewidth]{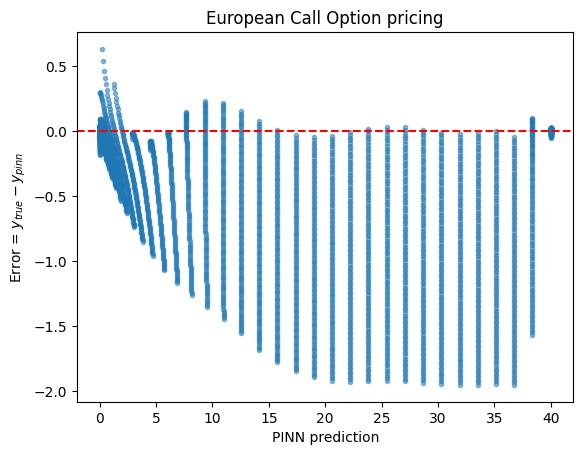}
    \includegraphics[width=0.45\linewidth]{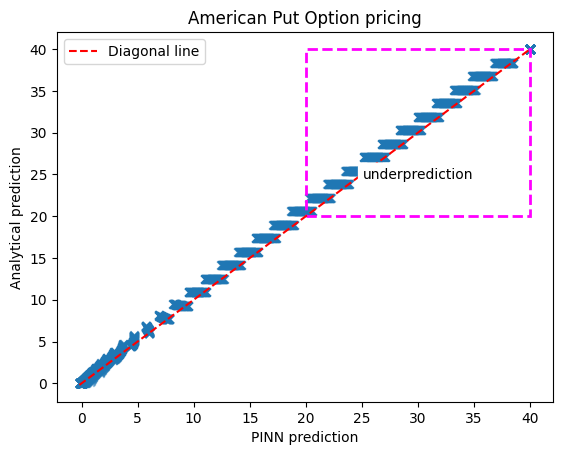}
    \caption{Error analysis for American put}
    \label{fig:american_put_err_analysis}
\end{figure}

Just like the European call, we conduct training for the American Put over 25,000 epochs. Once again, the PINN model demonstrates commendable accuracy in capturing the true solution's shape, as illustrated in \ref{fig:american_put_true_pred}. The performance metrics detailed in \ref{model_performance_metrics} further affirm the model's overall efficacy. However, upon closer examination, it becomes apparent that the PINN model consistently underestimates the true price, particularly in scenarios where the price is high, as depicted in \ref{fig:american_put_err_analysis}. This deviation is further elucidated in \ref{fig:american_put_err_analysis}.

\begin{figure}[h]
    \centering
    \includegraphics[width=0.8\linewidth]{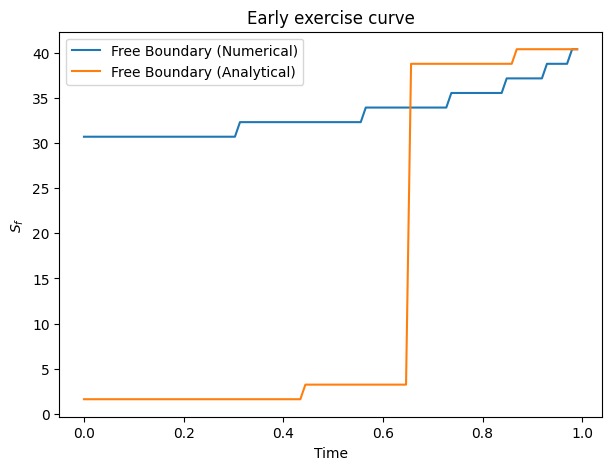}
    \caption{Exercise Boundary $S_f(t)$ for American Put}
    \label{fig:american_put_sf}
\end{figure}

An intriguing facet of the American Put lies in the concept of the exercise boundary. As previously elucidated, an American option can be exercised at any point, not solely at expiry. Interestingly, there exists a critical price $S_f(t)$, up to which it is advisable to promptly exercise the option. In this region the spot price is $K-S$. And in region (S>$S_f$) conventional BS model is applicable. The exercise boundary $S_f$ can thus be found as the $S_f(t) = max\{t | V = K - S\}$. The visualization of this exercise boundary is presented in \ref{fig:american_put_sf}. Notably, the PINN model exhibits a slight deviation in accurately identifying the true exercise boundary.

\subsection{Market Data}
% \todo[inline]{Ashish}

In addition to evaluating our Physics-Informed Neural Network (PINN) approach using historical market data, as elaborated in Section \ref{sec:data}, we aimed to extend our analysis to real-time market data. Our focus centers on call option on  Nasdaq 100 and put option on SPXC as the underlying assets. The PINN model for option pricing relies on two fixed inputs: volatility (\(\sigma\)) and risk-free interest rate (\(r\)). In our experimental setup, we utilize the implied volatility sourced from the Optionmetrics dataset \cite{wrds}. For the risk-free rate, we employ the average 1-year Treasury yield over the option period, acquired through web scraping from Yahoo Finance. Given these fixed parameters, we train the PINN model, with the required range of spot prices, separately for the two options. The separate models are then used at inference time to predict the market price.

The results of our approach on Market data are tabled in \ref{tab:market_results}. We can see that for both European call and American Put, while we do better then the analytical solution, but we still have errors when comparing to market price. There are multiple drivers possible for this, for ex, the market factors of supply and demand etc, or more basic problems like wrong choice of $\sigma$ and $r$. Having said that directionally the PINN approach does seem to work good, we see high corelation between the market and predicted price.

\begin{table}[ht]
  \centering
  \begin{tabular}{lcc}
    \toprule
    \textbf{Metric} & \textbf{NDX Eur. Call} & \textbf{SPXC Am. Put} \\
    \midrule
    RMSE($y_{bench}$, $y_{mkt}$) & 445.426 & 2.318 \\
    RMSE($y_{pinn}$, $y_{mkt}$) & \textbf{389.913} & \textbf{0.947 }\\
    cor($y_{pinn}$, $y_{mkt}$) & 0.630 & 0.730 \\
    \bottomrule
  \end{tabular}
  \caption{Performance metrics for NDX 100 European Call and SPXC American Put}
  \label{tab:market_results}
\end{table}

\begin{figure}
    \centering
    \includegraphics[width=0.45\linewidth]{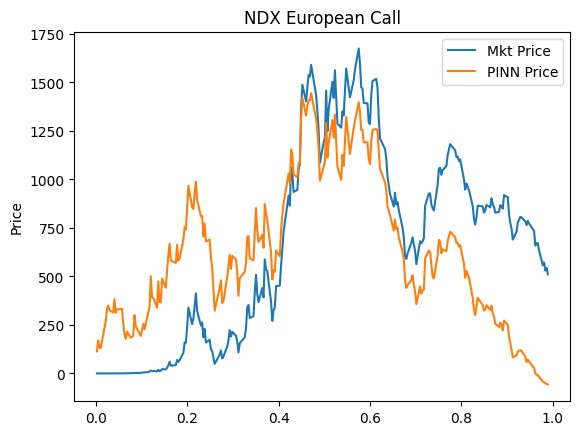}
    \includegraphics[width=0.45\linewidth]{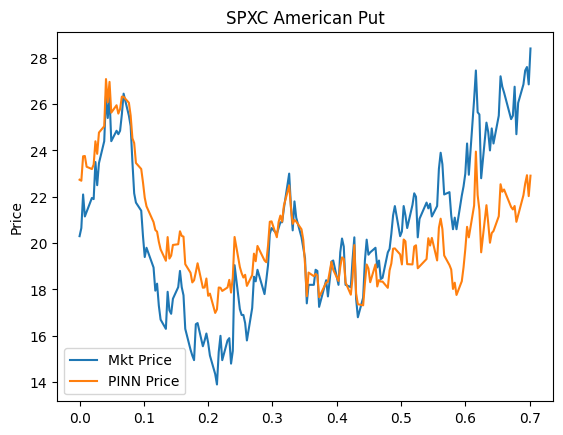}
    \caption{PINN results on NDX Call (left) and SPXC Put (right)}
    \label{fig:market_results}
\end{figure}

\section{Discussion}
% \todo[inline]{Ashish}

The application of the Physics-Informed Neural Network (PINN) model introduces a distinctive approach to option pricing. Through the implementation of PINN for both European and American options, we demonstrate its efficacy as a valuable tool in this domain. Our exploration extends to the development of a novel network architecture, drawing inspiration from the characteristic ramp shape of options. We posit that this architecture is particularly well-suited for enhancing the model's performance in option pricing tasks. Additionally, our investigation sheds light on the significance of penalty/sampling parameters and delineates approaches to multi-objective training for neural networks.

Having said that we our experiments reveal that PINNs do exhibit convergence challenges, particularly in specific regions of the domain. This observation prompts a deeper examination of the model's behavior and the identification of potential refinements. Additionally, both the PINN and Black-Scholes (BS) models necessitate careful consideration of various parameters, such as volatility and the risk-free rate. The selection of these parameters significantly influences model performance, adding an extra layer of complexity to the modeling process.

% \todo[inline]{Add prediction plot}

In conclusion, our study underscores the dual nature of PINNs, showcasing their effectiveness while also acknowledging challenges. The innovative insights gained from this exploration contribute to refining PINN applications in option pricing. As the field progresses, addressing convergence issues and optimizing parameter choices will play a crucial role in enhancing the robustness and reliability of PINN models in financial domains.

\section{Acknowledgements}
Help was taken from our coulleagues Anshit Verma, Jinsong Zheng and Pratyusha Pateel in helping prepare the manuscript of this paper.

% \section{Contribution Table}
% \begin{table}[ht]
%   \centering
%   \begin{tabular}{lc}
%     \toprule
%     \textbf{Task} & \textbf{Contributor(s)} \\
%     \midrule
%     PINN approach & Ashish, Pratyusha \\
%     Numerical benchmark & Anshit, Jinsong \\
%     Dataset & Ashish, Pratyusha, Anshit, Jinsong \\
%     Report + Poster & Ashish, Pratyusha, Anshit, Jinsong \\
%     \bottomrule
%   \end{tabular}
%   \caption{Task Contributions}
%   \label{table:task_contributions}
% \end{table}

% \begin{figure}
%     \centering
%     \includegraphics[width=0.7\linewidth]{assets/pinn_corelation.png}
%     \caption{Corelation b/w PINN and true solution.}
%     \label{fig:pinn_corealtion}
% \end{figure}

% \todo[inline]{Add co relation plot}

% \begin{figure}
%     \centering
%     \includegraphics[width=1\linewidth]{assets/pinn_pred.png}
%     \caption{Comparison of Analytical vs PINN prediction for European Call}
%     \label{fig:pinn_pred}
% \end{figure}

% \begin{table}
% \centering
% \begin{tabular}{lr}
% \hline
% \textbf{Loss Component} & \textbf{Value} \\
% \hline
% total loss & 7432.102681 \\
% loss (ivp+bvp) & 7332.079370 \\
% loss (pde) & 100.023314 \\
% \hline
% \end{tabular}
% \end{table}

% \todo[inline]{Add results: Ashish}

% \section{Next Steps}

% \begin{enumerate}
%     \item Debug the abnormality in the middle as shown in \ref{fig:pinn_pred}
%     \item Analyse effect of $\beta$
%     \item Experiment with LSTM cells in NN architecture
%     \item Replicate analysis for American Option
% \end{enumerate}

\newpage
{\small
\bibliographystyle{ieee_fullname}
% \newpage
\bibliography{egbib}
}

% \newpage

% \section{Project Milestone Video submission}
% \href{https://youtu.be/Ee9L5rb3oV0}{Midterm Project checkpoint video}
\end{document}